\documentclass[aps,prl,twocolumn,floats,amsmath,amssymb,showpacs]{revtex4}
\usepackage{bm}
\usepackage{graphicx}
\usepackage{dcolumn}

\begin{document}

\title{Universal anomalous diffusion of weakly damped particles}
\author{V. Bezuglyy$^{1,2}$,
M. Wilkinson$^{1}$ and B. Mehlig$^{2}$} \affiliation{
$^{1}$Department of Mathematics and Statistics, The Open University,
Walton Hall, Milton Keynes, MK7 6AA, England
\\$^{2}$Department of Physics,
G\"oteborg University, 41296 Gothenburg, Sweden
\\}

\begin{abstract}
We show that anomalous diffusion arises in two different models
for the motion of randomly forced and weakly damped particles: one
is a generalisation of the Ornstein-Uhlenbeck process with a
random force which depends on position as well as time, the other
is a generalisation of the Chandrasekhar-Rosenbluth model of
stellar dynamics, encompassing non-Coulombic potentials. We show
that both models exhibit anomalous diffusion of position $x$ and
momentum $p$ with the same exponents: $\langle x^2\rangle \sim C_x
t^2$ and $\langle p^2\rangle \sim C_p t^{2/5}$. We are able to
determine the prefactors $C_x$, $C_p$ analytically.
\end{abstract}

\maketitle

\section{Introduction}
\label{sec: 1}

In many systems the growth of a dynamical variable $X$ with time
$t$ satisfies $\langle X^2\rangle\sim C t^\alpha$ where the
angular brackets denote averaging. The process is said to be
anomalous diffusion if $\alpha\ne 1$. Anomalous diffusion may be a
consequence of a power-law built into the dynamical process, such
as in L\'evy flight models \cite{Met+00} , or it may be an \lq
emergent' property, where the anomalous exponent $\alpha$ is not a
direct consequence of power laws which are built into the model.
The latter case is more interesting, because non-integer 
exponents do not feature in the fundamental laws of physics, but there are
relatively few models where \lq emergent' anomalous diffusion can
be analysed exactly. In this paper we describe two physically
natural models for the diffusion of a particle which is
accelerated by random forces. If the damping is sufficiently weak
the particle can exhibit anomalous diffusion, having universal
exponents, the same for each model.

Our two models are generalisations of two classic models for
diffusion processes. The first is an extension of the
Ornstein-Uhlenbeck process \cite{Uhl+30}, in which a particle is
subjected to a rapidly fluctuating random force, and is damped by
viscous drag. In the generalised Ornstein-Uhlenbeck process the
random force depends upon the position of the particle and time
and is derived from a potential $\Phi(\mbox{\boldmath$x$},t)$. Earlier works
analysed this model in detail for one spatial dimension
\cite{Arv+05,Bez+06}. This model exhibits anomalous diffusion in one
dimension. Here we discuss higher spatial dimensions, where
the mechanism for anomalous diffusion is significantly different,
as was suggested (for a closely related model) in
\cite{Gol+91,Ros92}. Here we obtain exact formulae for the momentum
distribution of the generalised Ornstein-Uhlenbeck process in two
and three dimensions (for a particle which is initially at rest).
We use these to obtain precise asymptotic formulae for the growth
of the second moment of the coordinate: the second moments scale
as $\langle p^2\rangle\sim t^{2/5}$ and $\langle x^2\rangle\sim
t^2$ in the anomalous diffusion regime (the exponents are the same
as those obtained in \cite{Ros92}; we obtain the prefactor
exactly).

We also discuss an extension of the Chandrasekhar-Rosenbluth model
for diffusion \cite{Cha43,Ros+57}, in which a test particle
interacts with a gas of point masses via a pair potential. The
interaction should cause small changes of momentum, which can be
modelled as a diffusion process. Usually the interaction is
gravitational, and the application is to the motion of stars in
galaxies, but here we simplify the problem by considering a
non-singular weak interaction potential. We show that,
surprisingly, the diffusion tensor has the same form as for the
generalised Ornstein-Uhlenbeck process, and that consequently
there is anomalous diffusion with the same universal exponents.
For this model too, we derive diffusion coefficients for this
model precisely in terms of the microscopic parameters.

The anomalous diffusion effect which we describe is analysed 
by introducing a diffusion process describing the fluctuations
of the momentum $p$ of a particle in response to a spatially
and temporally fluctuating random potential. The diffusion coefficient
of this process, $D(p)$,  is a function of the momentum of the particle.
We remark that this approach to formulating the equation of motion 
was first introduced by Sturrock \cite{Stu66}, and that similar
developments appeared later in the mathematical literature
(see \cite{Agu+09} and references cited therein). Our paper 
is the first work to give a solution to the generalised 
Ornstein-Uhlenbeck process in two or three dimensions.  

\section{Generalised Ornstein-Uhlenbeck model}
\label{sec: 2}

We consider a particle of mass $m$ with momentum $\mbox{\boldmath$p$}$
subjected to the generalised Ornstein-Uhlenbeck process
\cite{Bez+06}. This is described by equations of motion
\begin{equation}
\label{eq: 2.1} 
\dot{\mbox{\boldmath$x$}}=\frac{\mbox{\boldmath$p$}}{m}\ ,\ \ \ 
\dot{\mbox{\boldmath$p$}}=-\gamma \mbox{\boldmath$p$} + \mbox{\boldmath$f$}(\mbox{\boldmath$x$},t)
\end{equation}
where $\mbox{\boldmath$x$}$ is the particle's position. The particle
experiences two types of forces: a drag force $-\gamma \mbox{\boldmath$p$}$
with $\gamma$ being a damping rate and a random force
$\mbox{\boldmath$f$}(\mbox{\boldmath$x$},t)$. Unlike the classic Ornstein-Uhlenbeck
process, where the force depends only upon time, in our 
generalised model the random force depends upon the position as
well. We assume that $\mbox{\boldmath$f$}(\mbox{\boldmath$x$},t)$ is a force 
derived from a random potential varying in time and space, i.e.
$\mbox{\boldmath$f$}(\mbox{\boldmath$x$},t)=-\nabla \Phi (\mbox{\boldmath$x$},t)$, where $\Phi
(\mbox{\boldmath$x$},t)$ has statistics
\begin{equation}
\label{eq: 2.2} \langle \Phi (\mbox{\boldmath$x$},t) \rangle = 0\ ,\ \langle
\Phi (\mbox{\boldmath$x$},t)\Phi (\mbox{\boldmath$x$}',t')\rangle =
C(|\mbox{\boldmath$x$}-\mbox{\boldmath$x$}'|,|t-t'|)\ .
\end{equation}
The correlation function $C(x,t)$ has temporal and spatial scales,
$\tau$ and $\xi$, respectively. We consider the case where
$\gamma\tau\ll 1$, implying that the momentum satisfies a
diffusion equation. We define a momentum scale $p_0=m\xi/\tau$,
such that if $|\mbox{\boldmath$p$}|\gg p_0$ the force experienced by the
particle decorrelates much more rapidly than the force experienced
by a stationary particle. We consider the limit where $\vert
\mbox{\boldmath$p$}\vert\gg p_0$, which is realised for weak damping. The
dynamics of (\ref{eq: 2.1}) can be described by a diffusion equation
for the probability density of the momentum, $P(\mbox{\boldmath$p$},t)$: we
now consider how to derive this diffusion equation.

The dynamics of the momentum can be approximated by a Langevin
process. Small increments of components $p_i$ of the 
momentum vector $\mbox{\boldmath$p$}$ 
may be written as
\begin{equation}
\label{eq: 2.3} \delta p_i=-\gamma p_i \delta t + \delta w_i
\end{equation}
where $\delta w_i$ is the impulse exerted by the $i$-th component
of the force $\mbox{\boldmath$f$}$ on the particle in time $\delta t$,
\begin{eqnarray}
\label{eq: 2.4} \delta w_i(t_0)&=&\int _{t_0}^{t_0+\delta t} {\rm d}t_1 \
f_i(\mbox{\boldmath$x$}(t_1),t_1) \nonumber\\
&=& \int _{t_0}^{t_0+\delta t} {\rm d}t_1 \
f_i(\mbox{\boldmath$p$}t_1/m,t_1)+O(\delta t^2)\ .
\end{eqnarray}
The Langevin process \eqref{eq: 2.3} is equivalent to the
Fokker-Planck equation describing time evolution of the
probability density of momentum $P(\mbox{\boldmath$p$},t)$. In order to
construct the equation we need to know drift and diffusion
coefficients, $v_i=\langle \delta p_i \rangle/\delta t$ and
$D_{ij}=\langle \delta p_i \delta p_j \rangle/2\delta t$
respectively. Using the definition of the increment $\delta w_i$
we find
\begin{eqnarray}
\label{eq: 2.5}
\langle \delta w_i \delta w_j \rangle &\sim & \int _0
^{\delta t} {\rm d}t_1 \int _0 ^{\delta t} {\rm d}t_2 \langle
f_i(\mbox{\boldmath$p$}t_1/m,t)f_j(\mbox{\boldmath$p$}t_2/m,t_2) \rangle \nonumber \\
&\sim & \delta t \int _{-\infty} ^{\infty} {\rm d}t_1 \ \langle
f_i(\mbox{\boldmath$0$},0)f_j(\mbox{\boldmath$p$}t_1/m,t_1)\rangle \nonumber \\
&=&2D_{ij}\delta t
\end{eqnarray}
(the second step is justified when $\delta t$ is large compared to
$\tau$ but small compared to $\gamma^{-1}$).

The components $D_{ij}$ of the momentum diffusion tensor ${\bf D}$
depend upon the direction of the momentum. For the case where the
momentum is aligned with the $x$-axis (that is, where
$\mbox{\boldmath$p$}=p{\bf e}_1$, the coefficients of the diffusion matrix
are expressed in terms of the correlation function of the
potential as follows:
\begin{eqnarray}
\label{eq: 2.6} D_{xx}&=&-\frac{m}{2p}\int_{-\infty}^\infty {\rm
d}R\ \frac{\partial ^2 C}{\partial R^2}(R,mR/p)\nonumber
\\
D_{yy}&=&-\frac{m}{2p}\int_{-\infty}^\infty {\rm d}R\
\frac{1}{|R|}\frac{\partial C}{\partial R}(R,mR/p)\nonumber \\
D_{xy}&=&0\ .
\end{eqnarray}
(In three dimensions $D_{zz}=D_{yy}$ and $D_{xz}=D_{yz}=0$.) For
other directions of the momentum, the elements of the momentum
diffusion tensor can be obtained by applying a rotation matrix. If
${\bf O}$ is a rotation matrix which rotates the momentum vector
$\mbox{\boldmath$p$}$ into $\mbox{\boldmath$p$}'={\bf O}\mbox{\boldmath$p$}=p{\bf e}_1$ which is
aligned with the $x$-axis, then the elements $D'_{ij}$ of the
diffusion matrix in the transformed coordinate system are given by
(\ref{eq: 2.6}). The diffusion matrix in the original coordinate
system is ${\bf D}={\bf O}\,{\bf D}'\,{\bf O}^{\rm T}$.

Having considered the diffusive fluctuations of the momentum, we
now consider its drift, $\langle \delta p_i \rangle = -\gamma p_i
\delta t+\langle \delta w_i \rangle$. Expanding $f_i(\mbox{\boldmath$x$},t)$
we obtain (in $d$ dimensions)
\begin{equation}
\label{eq: 2.7} 
f_i(\mbox{\boldmath$x$},t)=f_i(\mbox{\boldmath$0$},t)+\sum _{j=1}^{d}
\frac{\partial f_i(\mbox{\boldmath$0$},t)}{\partial x_j}x_j(t)
\end{equation}
where $x_j(t)$ can be written as a solution of (\ref{eq: 2.1}),
\begin{eqnarray}
\label{eq: 2.8} x_j(t) &=& \frac{1}{m} \int _0^t {\rm d}t_1 \int _0^{t_1}
{\rm d}t_2 \ {\rm exp}[-\gamma(t_1-t_2)] \nonumber \\ &\times&
f_j(\mbox{\boldmath$p$}t_2/m,t_2)\ .
\end{eqnarray}
Combining together (\ref{eq: 2.4}), (\ref{eq: 2.7}) and (\ref{eq: 2.8})
we obtain
\begin{eqnarray}
\label{eq: 2.9} 
\langle \delta w_i \rangle  &\sim & \frac{1}{m}\sum _{j=1} ^d
\int _0^{\delta t} {\rm d}t_1 \int _0^{t_1} {\rm d}t_2 \int _0^{t_2}{\rm d}t_3 \
\exp[-\gamma (t_2-t_3)] \nonumber \\
&&\times \bigg \langle \frac{\partial f_i(\mbox{\boldmath$0$},t_1)}{\partial
x_j} f_j(\mbox{\boldmath$p$}t_3/m,t_3) \bigg \rangle
\ .
\end{eqnarray}
We approximate ${\rm exp}[-\gamma (t_2-t_3)]$ by unity for $\gamma
\tau \ll 1$, and use the assumption that $\delta t\gg \tau$ to obtain
\begin{eqnarray}
\label{eq: 2.10} 
\langle \delta w_i \rangle & \sim & \frac{\delta
t}{2m} \sum _{j=1}^d \int _{-\infty}^{\infty} {\rm d}t \ t \bigg \langle
\frac{\partial f_i(\mbox{\boldmath$0$},0)}{\partial x_j}
f_j(\mbox{\boldmath$p$}t/m,t) \bigg \rangle \nonumber \\
&=& \delta t \sum _{j=1}^d \frac{\partial}{\partial
p_j}D_{ij}(\mbox{\boldmath$p$})\ .
\end{eqnarray}
Taking account of the fact that the drift coefficients contain
derivatives of the diffusion coefficients, we obtain the
Fokker-Planck equation,
\begin{equation}
\label{eq: 2.11} \frac{\partial P}{\partial
t}=\frac{\partial}{\partial \mbox{\boldmath$p$}}\cdot \left( \gamma \mbox{\boldmath$p$}
+ {\bf D}(\mbox{\boldmath$p$}) \frac{\partial}{\partial \mbox{\boldmath$p$}} \right) P\ .
\end{equation}
We are primarily interested in the case of weak damping, where the
typical momentum satisfies $\vert \mbox{\boldmath$p$}\vert\gg p_0$. In this
limit, the diffusion coefficients have an algebraic dependence
upon $p=\vert\mbox{\boldmath$p$}\vert$. For diffusion of the momentum vector
parallel to its direction, when $p \gg p_0$ the leading term is of
order $p^{-3}$ and the result is given by
\begin{equation}
\label{eq: 2.12} D_{xx}\sim\frac{D_3 p_0^3}{p^3}, \;\;\;
D_3=-\frac{m^3}{4 p_0^3}\int _{-\infty}^{\infty} {\rm d}R R^2
\frac{\partial ^4 C}{\partial R^2\partial t^2}(R,0)\ .
\end{equation}
For diffusion of $\mbox{\boldmath$p$}$ in a direction perpendicular to its
direction, we have
\begin{equation}
\label{eq: 2.13} 
D_{yy}\sim\frac{D_1 p_0}{p}, \;\;\;
D_1=-\frac{m}{2p_0}\int _{-\infty}^{\infty} {\rm d}R\ \frac{1}{| R|}
\frac{\partial  C}{\partial R}(R,0)\ .
\end{equation}
Note that when $p\gg p_0$, the diffusion of the direction of the
momentum vector is much more rapid than diffusion of its
magnitude. This makes the behaviour of the generalised
Ornstein-Uhlenbeck process in two or more dimensions very
different from its behaviour in one dimension.

Results equivalent to (\ref{eq: 2.12}) and (\ref{eq: 2.13}) were
obtained in \cite{Gol+91,Ros92} in an analysis of a closely related
model.

\section{Generalised Chandrasekhar-Rosenbluth model}
\label{sec: 3}

In this section we consider motion of a particle travelling
through an infinite homogeneous population of background
particles. We assume that the test particle interacts 
with the background particles, and the background
particles do not interact with each other. As the test particle
moves, its interaction with each of the background particles
causes small changes of its velocity. When the number of the
background particles is very large, the velocity of the test
particle changes rapidly and in an unpredictable way, so that its
motion can be described by a diffusion process. The
test particle could be a star moving in a galaxy interacting with
the background stars (the interaction between the background stars
is not considered), so that the force of interaction is
gravitational and thus proportional to the inverse square of the
distance $r$ between stars. This problem was originally studied 
by Chandrasekhar \cite{Cha43}, who found that the
test particle experiences a gradual decrease of the velocity in
the direction of motion. This phenomenon is called \lq dynamical
friction'. For \lq slow' particles (with velocities much smaller than 
some representative velocity scale) this deceleration is proportional to
the velocity of the particle $v$ (analogous to the Stokes's law
for a drag force for a particle is a viscous medium). For sufficiently 
\lq fast' particles the deceleration is proportional
to $v^{-2}$.

Subsequently, Rosenbluth {\sl et al.} \cite{Ros+57} studied the 
diffusion of momentum of the test particle in this model in greater
depth. They found that in the \lq fast' regime the diffusion coefficient 
of the momentum in the direction parallel to the direction of motion is proportional 
to $v^{-3}$, while the diffusion coefficients in the plane perpendicular to 
the direction of motion is proportional to $v^{-1}$. These dependences 
are equivalent to the momentum dependences of the radial and transverse 
fluctuations of the momentum in the generalised Ornstein-Uhlenbeck model,
obtained in equations (\ref{eq: 2.12}) and (\ref{eq: 2.13}). The expressions 
for these diffusion coefficients obtained in \cite{Ros+57} contain logarithmic
terms due to the long-ranged nature of the Coulombic potential, which makes
it difficult to write down precise formulae. In order to illuminate
the relation between the generalised Ornstein-Uhlenbeck model and the 
Chandrasekhar-Rosenbluth model in the simplest context, in this section 
we consider the latter model for the case when the interaction
is described by a short-range potential $U(r)$ of some rather
general radially symmetric form. We obtain precise expressions for these 
diffusion coefficients and show 
that the results for the scaling of the diffusion
coefficients are the same as in the generalised Ornstein-Uhlenbeck
process. This leads to the same anomalous diffusion behaviour in
both models.

We shall only discuss the two-dimensional case for simplicity and
proceed as follows (some of the presentation adapts the discussion
of the Chandrasekhar model in \cite{Bin+94}).  
We first calculate the change of the velocity
of a test particle due to the encounter with a stationary
background particle. We denote components of this change by
$\Delta v'_{\parallel}$ and $\Delta v'_{\perp}$, in the directions
parallel and perpendicular to the initial velocity of the test
particle. Next, we consider  the change of the velocity of the
test particle for the case when the background particle propagates
with velocity $\mbox{\boldmath$v$}_b$. We denote components of this change by
$\Delta v_{\parallel}$ and $\Delta v_{\perp}$. Using geometrical
arguments, we then express $\Delta v_{\perp}$ and $\Delta
v_{\parallel}$ in terms of $\Delta v'_{\perp}$ and $\Delta
v'_{\parallel}$.

\begin{figure}[t]
\centerline{\includegraphics[width=8.0cm]{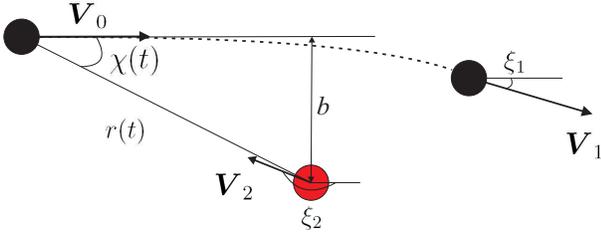}}
\caption{\label{fig: 1} 
The encounter between a test
particle (black) and a background particle (shaded, red online). The interaction
causes a small change of the velocity of both particles.}
\end{figure}

Let us consider a test particle of mass $m$ moving in the
horizontal direction with the initial velocity
$\mbox{\boldmath$V$}_0=(V_0,0)$. The test particle interacts with a
background particle of mass $M$, which is initially at rest. After
an encounter the test particle propagates with velocity
$\mbox{\boldmath$V$}_1$ described by its magnitude $V_1$ and the polar angle
$\xi_1$, and the background particle moves with velocity
$\mbox{\boldmath$V$}_2$ described by its magnitude $V_2$ and the polar angle
$\xi_2$ (see Fig.~\ref{fig: 1}).

The changes of the velocity of the test particle in the direction
parallel and perpendicular to $\mbox{\boldmath$V$}_0$ are
\begin{align}
\label{eq: 3.1}
\Delta v'_{\parallel}&=V_1\cos \xi_1 - V_0, \nonumber\\
\Delta v'_{\perp}&=V_1\sin \xi_1\ .
\end{align}
The conservation of momentum before and after the encounter yields
\begin{eqnarray}
\label{eq: 3.2}
mV_0&=&mV_1\cos\xi_1 + MV_2\cos \xi_2
\nonumber\\
&0&=mV_1\sin\xi_1 + MV_2\sin \xi_2
\end{eqnarray}
and from the conservation of energy we obtain
\begin{equation}
\label{eq: 3.3}
mV_0^2=mV_1^2+MV_2^2\ .
\end{equation}
This enables us to write an equation for $V_1$:
\begin{equation}
\label{eq: 3.4}
V_1^2(m+M)-2mV_0V_1\cos\xi_1 +V_0^2(m-M)=0\ .
\end{equation}
We assume that the encounter induces only a small change of the
direction of motion of the test particle, so that $\xi_1$ can be
taken as being small. In this approximation the solution of
equation (\ref{eq: 3.4}) is
\begin{equation}
\label{eq: 3.5} 
V_1=V_0(1-\alpha\xi_1^2)+O(\xi_1^4)
\end{equation}
where $\alpha=m/(2M)$. We substitute $V_1$ from (\ref{eq: 3.5}) into equation (\ref{eq: 3.1})
and obtain
\begin{eqnarray}
\label{eq: 3.6}
\Delta v'_{\parallel}&\sim&-V_0\beta\xi_1^2\nonumber\\
\Delta v'_{\perp}&\sim& V_0\xi_1
\end{eqnarray}
where $\beta=\alpha+1/2=(m+M)/2M$. In the small-angle
approximation $\xi_1$ is defined by the change of the momentum in
the perpendicular direction, so that $\xi_1\sim \Delta
p_{\perp}/(mV_0)$. The change of the momentum $\Delta p_{\perp}$
is determined by the force of interaction between particles 
separated by distance $r$ with magnitude
$f(r)=-{\rm d}U(r)/{\rm d}r$, so that $\Delta p_{\perp}$ can be written as
\begin{equation}
\label{eq: 3.7} 
\Delta p _{\perp} = -\int _{-\infty}^{\infty} {\rm d}t
\frac{{\rm d}U}{{\rm d}r}[r(t)] \sin \chi(t)
\end{equation}
where $r(t)$ is a distance between the particles and $\chi(t)$ is
an angle between $\mbox{\boldmath$V$}_0$ and the vector connecting the
particles. We define an impact parameter $b$ to be the initial distance
between the test and background particles along the axis
perpendicular to $\mbox{\boldmath$V$}_0$. From figure~\ref{fig: 1} we find
that $r(t) = \sqrt{x(t)^2+b^2}$ and $\sin \chi(t) =
b/\sqrt{x(t)^2+b^2}$, where $x(t)$ is a coordinate of the test
particle along the direction parallel to $\mbox{\boldmath$V$}_0$. Changing
the variable $x(t)=V_0 t$, in the weak-scattering limit where the deflection
is small we obtain
\begin{equation}
\label{eq: 3.8}
\Delta p _{\perp} \approx -\frac{1}{V_0} \int _{-\infty}^{\infty}
{\rm d}x \ \frac{{\rm d}U}{{\rm d} r}(\sqrt{x^2+b^2})
\frac{b}{\sqrt{x^2+b^2}}\ .
\end{equation}
If we denote an integral
\begin{equation}
\label{eq: 3.9}
I(b)=-\int _{-\infty}^{\infty} {\rm d}x \ \frac{{\rm d}U}{{\rm d}r} (\sqrt{x^2+b^2})\frac{b}{\sqrt{b^2+x^2}}
\end{equation}
we obtain
\begin{equation}
\label{eq: 3.10} 
\xi_1(b,V_0)\sim \frac{I(b)}{mV_0^2}\ .
\end{equation}
Using this relation we find
\begin{eqnarray}
\label{eq: 3.11}
\Delta v'_{\parallel}&=& -\frac{\beta I^2(b)}{m^2V_0^3}\nonumber \\
\Delta v'_{\perp}&=&\frac{I(b)}{mV_0}
\ .
\end{eqnarray}
Thus, the contribution to the change of the velocity of the test
particle due to a single encounter is proportional to $V_0^{-3}$
and $V_0^{-1}$ in the directions parallel and perpendicular to
$\mbox{\boldmath$V$}_0$, respectively. Averaging over collisions 
with many particles is expected to add another factor of $V_0$, as 
the particle propagates with this velocity (see also the derivation below).
This suggests that the second moments of the change of the
velocity scale as $V_0^{-5}$ and $V_0^{-1}$ in the directions
parallel and perpendicular to $\mbox{\boldmath$V$}_0$, respectively. While
the latter result is consistent with the behaviour of the
diffusion coefficient in the generalised Ornstein-Uhlenbeck
process, the former result is different, as in the previous model
the result is $\langle \Delta v^2_{\parallel}\rangle \sim v^{-3}$.
However, if the background particles are not stationary, these
estimates must be corrected, as shown below.

\begin{figure}[t]
\centerline{\includegraphics[width=8.0cm]{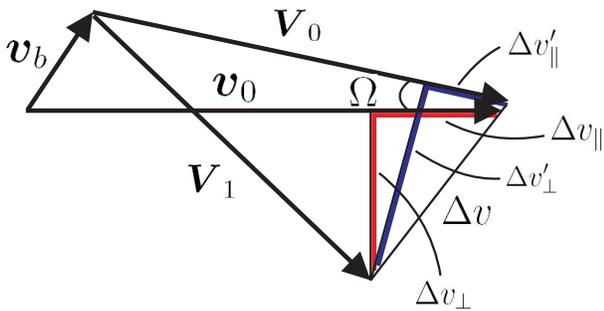}}
\caption{\label{fig: 2} 
Geometrical construction illustrating the changes of the velocity of the test
particle $\Delta v$ parallel and perpendicular to the relative
velocity $\mbox{\boldmath$V$}_0$ ($\Delta v'_{\parallel}$ and $\Delta
v'_{\perp}$, solid lines, blue online) and the velocity of the test
particle $\mbox{\boldmath$v$}_0$ ($\Delta v_{\parallel}$ and $\Delta
v_{\perp}$, solid lines, red online).}
\end{figure}

We assume that the background particle moves with velocity
$\mbox{\boldmath$v$}_b$, in which case the discussion above is valid if
$\mbox{\boldmath$V$}_0$ is a relative velocity of the test particle in the
frame of reference moving with the background particle. The
velocity of the test particle in a fixed frame of reference is
therefore $\mbox{\boldmath$v$}_0=\mbox{\boldmath$V$}_0+\mbox{\boldmath$v$}_b$. We are interested in
the changes of the velocity of the test particle in the directions
parallel and perpendicular to $\mbox{\boldmath$v$}_0$. We denote these
$\Delta v_{\parallel}$ and $\Delta v_{\perp}$ and deduce from
figure~\ref{fig: 2}:
\begin{eqnarray}
\label{eq: 3.12}
\Delta v_{\parallel}&=&\Delta v'_{\parallel} \cos \Omega+\Delta
v'_{\perp}\sin \Omega, \nonumber \\
\Delta v_{\perp}&=&\Delta v'_{\perp}\cos \Omega-\Delta
v'_{\parallel}\sin \Omega
\end{eqnarray}
where $\Omega$ is an angle between $\mbox{\boldmath$V$}_0$ and $\mbox{\boldmath$v$}_0$. These relations 
are equivalent to the rotation of the coordinate
system by angle $\Omega$. Assuming that $v_0\gg v_b$, where
$v_b=|\mbox{\boldmath$v$}_b|$ and $v_0=|\mbox{\boldmath$v$}_0|$, 
we obtain 
\begin{equation}
\label{eq: 3.13}
\Omega \sim \frac{\mbox{\boldmath$v$}_b\wedge \mbox{\boldmath$V$}_0}{V_0^2}
\equiv \frac{v_{b\perp}}{V_0}
\end{equation} 
 and $v_0\sim V_0$.
In this approximation, to the leading order in $v_0$, we
have
\begin{eqnarray}
\label{eq: 3.14}
\Delta v_{\parallel} &\approx& \frac{I(b)v_{b\perp}}{mv_0^2} 
\nonumber\\
\Delta v_{\perp} &\approx& \frac{I(b)}{mv_0}
\ .
\end{eqnarray}

We now imagine that the test particle is travelling through an
infinite homogenous population of the background particles with
the spatial density number $n$ (measuring a number of particles
per unit area) and the probability density of the velocity is 
$f(\mbox{\boldmath$v$}_b)$. Let ${\rm d}N$ be the number of background 
particles it encounters in time $\Delta t$ with velocity $\mbox{\boldmath$v$}_b$ 
in a volume element of velocity space ${\rm d}\mbox{\boldmath$v$}_b$ 
and impact parameter between $b$ and $b+{\rm d}b$. This is the number of 
particles in two thin stripes, each of width ${\rm d}b$ and length equal to the distance
travelled by the particle in $\Delta t$, multiply by the probability
$f(\mbox{\boldmath$v$}_b)\,{\rm d}\mbox{\boldmath$v$}_b$. We have
\begin{equation}
\label{eq: 3.15}
{\rm d} N \sim 2 n V_0 \Delta t \, {\rm d}b \times f(\mbox{\boldmath$v$}_b)\,{\rm d}\mbox{\boldmath$v$}_b
\ .
\end{equation}
In order to obtain the total contribution of many background
particles with different impact parameters and velocities, we
integrate over $b$ and $\mbox{\boldmath$v$}_b$,
\begin{eqnarray}
\label{eq: 3.16}
\frac{\langle \Delta v_{\parallel}^2 \rangle}{2\Delta t} 
&=& \frac{n}{m^2v_0^3} \int_0^\infty {\rm d}b \ I^2(b)\int {\rm d}\mbox{\boldmath$v$}_b\ f(\mbox{\boldmath$v$}_b) |\mbox{\boldmath$v$}_{b \perp}|^2 
\nonumber \\
\frac{\langle \Delta v_{\perp}^2 \rangle}{2\Delta t} &=&
\frac{ n}{m^2v_0} \int_0^\infty {\rm d}b \ I^2(b)
\ .
\end{eqnarray}
Here we used the assumption that $U(r)$ is a short-ranged, allowing us
to let the upper limit of the integral over $b$ approach infinity. 
In the case of the original Chandrasekhar-Rosenbluth model, 
an upper limit to the impact parameter must be introduced 
because of the long-range interaction between the particles. 
This leads to logarithmic correction terms \cite{Cha43,Ros+57}.

Equations (\ref{eq: 3.16}), describing the velocity increments for the 
Chandrasekhar-Rosenbluth 
model has the same scaling (as a function of
$v_0$) as scaling of the diffusion coefficients for the generalised 
Ornstein-Uhlenbeck model (as a function of $p$; see equations 
(\ref{eq: 2.12}) and (\ref{eq: 2.13})). This indicates that
the anomalous diffusion behaviour of these models is equivalent.

\section{Probability density function and moments of the momentum}
\label{sec: 4}

Now we return to the generalised Ornstein-Uhlenbeck process and obtain
the closed-form solution of the Fokker-Planck equation for a
particular choice of the initial conditions. We use this solution
to obtain an exact expression for the growth of the moments of the
momentum.

We first consider the two-dimensional case. The probability density for the 
momentum satisfies equation (\ref{eq: 2.11}), with the diffusion coefficients
given by (\ref{eq: 2.12}) and (\ref{eq: 2.13}). We transform to polar coordinates
and seek a probability density $P(p,\theta,t)$, and consider the case when 
the particle is initially at rest, so that the initial condition is 
$P(p,\theta,0)=\delta(p)$. This circularly symmetric solution, $P=\rho(p,t)$, satisfies
\begin{equation}
\label{eq: 4.1} 
\frac{\partial \rho}{\partial
t}=\frac{D_3p_0^3}{p^3}\frac{\partial^2 \rho}{\partial
p^2}+\left(\gamma p -\frac{2D_3p_0^3}{p^4}\right)\frac{\partial
\rho}{\partial p} + 2\gamma \rho
\ .
\end{equation}
By analogy with the solution of the one-dimensional generalised Ornstein-Uhlenbeck
model, we find the following normalised closed-form solution of (\ref{eq: 4.1}):
\begin{eqnarray}
\label{eq: 4.2} 
\rho(p,t)&=& \frac{5}{\Gamma(2/5)}
\frac{\gamma^{2/5}}
{[5D_3p_0^3(1-{\rm e}^{-5\gamma t})]^{2/5}}\nonumber \\
&\times& {\rm exp}\left[-\frac{\gamma p^5}{5D_3p_0^3(1-{\rm
e}^{-5\gamma t})} \right]\ .
\end{eqnarray}
In the long-time limit the density is non-Maxwellian given by
\begin{equation}
\label{eq: 4.3} 
\rho_0(p)= \frac{5\gamma^{2/5}}
{\Gamma(2/5)(5D_3p_0^3)^{2/5}} {\rm exp}\left[-\frac{\gamma
p^5}{5D_3p_0^3} \right]\ .
\end{equation}
Using the probability density (\ref{eq: 4.2}) we determine the
$l$-th moment of $p$,
\begin{eqnarray}
\label{eq: 4.4} 
\langle p^l(t)\rangle &=& \int _0^{\infty} {\rm d}p \ p^{l+1}
\rho(p,t) \nonumber \\
&=& \left( \frac{5D_3p_0^3}{\gamma}\right)^{l/5}
\frac{\Gamma[(2+l)/5]}{\Gamma(2/5)}(1-{\rm e}^{-5\gamma t})^{l/5}
\ .
\end{eqnarray}
We remark that an additional factor of $p$ in the expression above
appears as a weight in the transformation to polar coordinates.
\begin{figure}[t]
\centerline{\includegraphics[width=9.0cm]{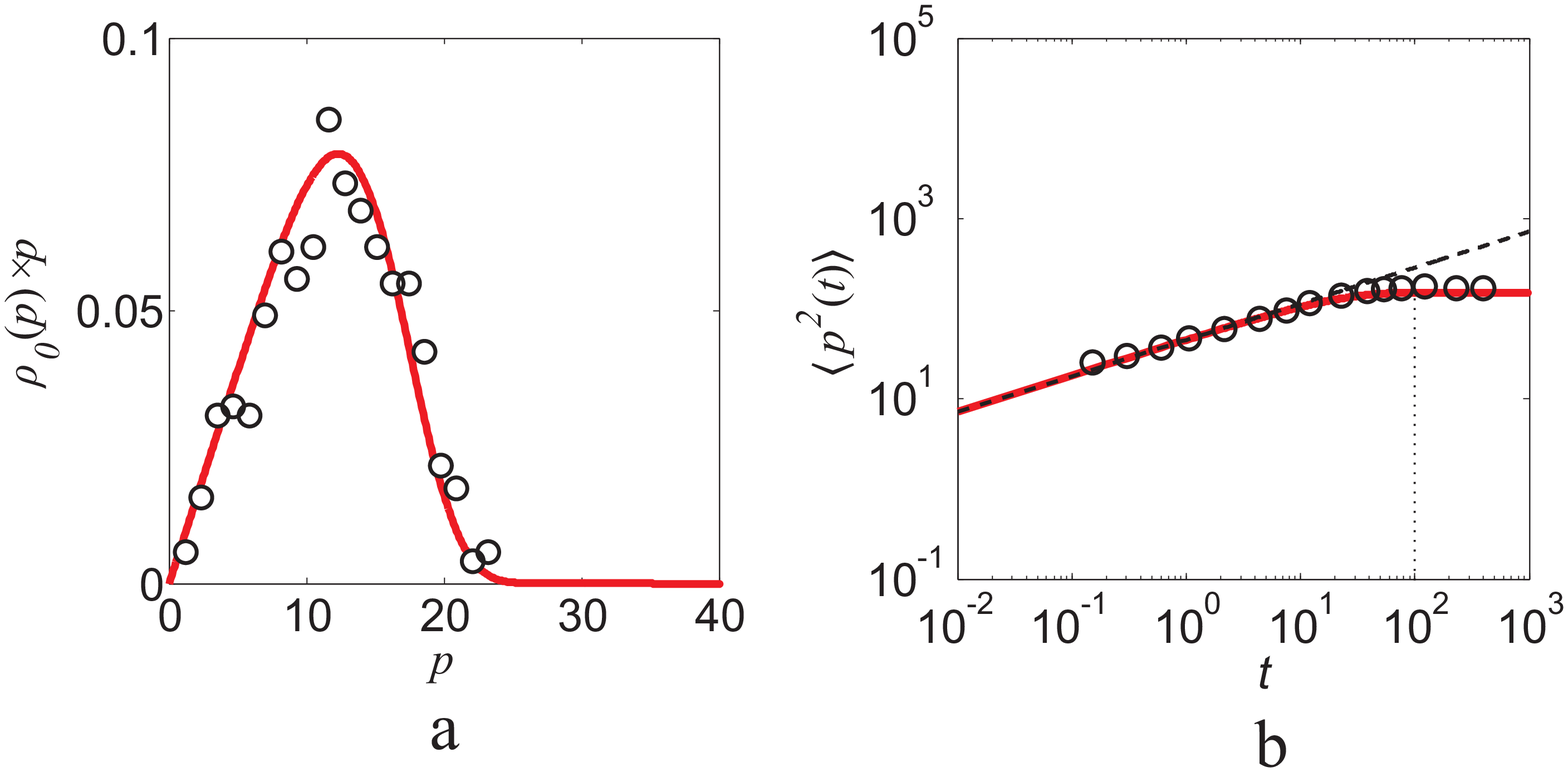}}
\caption{\label{fig: 3} Shows results of the
numerical simulation of equation~(\ref{eq: 2.1}) for the motion in the
two-dimensional potential force field. Panel {\bf a} shows
stationary non-Maxwellian density function (\ref{eq: 4.3}) (solid line) and
data from the numerical simulation (circles). Panel {\bf b} shows
the second moment of the momentum (\ref{eq: 4.4}) with $l=2$ (solid
line) and data from the numerical simulation (circles). The dashed
line shows the slope $t^{2/5}$ and dotted line indicates time
$\gamma^{-1}$ at which the density becomes stationary. The results
are for the case of a Gaussian correlation function of the potential
$C(x,t)=\sigma^2{\rm exp}[-x^2/(2\eta^2)-t^2/(2\tau^2)]$ with
$\sigma=15$, $\tau=0.1$ and $\eta=0.1$. The other parameters were
$m=1$ and $\gamma=0.01$.}
\end{figure}

In the three-dimensional case we find a similar solution to 
equation~(\ref{eq: 2.10}) in the case where the particles are 
initially stationary. We write this equation in spherical 
polar coordinates, and seek a spherically 
symmetric solution, $P(p,\theta,\phi,t)=\rho(p,t)$. The solution of the 
corresponding equation for $\rho(p,t)$ is obtained similarly to the
two-dimensional case and we have
\begin{eqnarray}
\label{eq: 4.5}
\rho(p,t)&=&\frac{5}{\Gamma(3/5)}\frac{\gamma^{3/5}}{[5D_3p_0^3
(1-{\rm e}^{-5\gamma t})]^{3/5}} \nonumber \\ 
&\times& {\rm
exp}\left[ -\frac{\gamma p^5}{5D_3p_0^3(1-{\rm e}^{-5\gamma t})}
\right].
\end{eqnarray}
This determines moments of the momentum
\begin{eqnarray}
\label{eq: 4.6}
\langle p^l(t)\rangle &=& \left(
\frac{5D_3p_0^3}{\gamma}\right)^{l/5}
\frac{\Gamma[(3+l)/5]}{\Gamma(3/5)} \nonumber \\
&\times& (1-{\rm e}^{-5\gamma t})^{l/5}\ .
\end{eqnarray}
For both two- and three-dimensional cases we obtain that at short
times the variance of the momentum grows as
\begin{equation}
\label{eq: 4.7}
\langle p^2(t)\rangle \sim t^{2/5}
\ .
\end{equation}
Thus, at short times the momentum diffuses anomalously with the
same exponent as in the one-dimensional model \cite{Gol+91,Ros92}. 
The results for the stationary probability
density and diffusion of the momentum in the two-dimensional case
were verified by a numerical simulation, documented in figure~\ref{fig: 3}.

\section{Spatial diffusion}
\label{sec: 5}

In this section we find the mean-square value of the displacement
of a particle which starts at the origin:
\begin{eqnarray}
\label{eq: 5.1}
\langle |\mbox{\boldmath$x$}(t)|^2\rangle &=& \frac{1}{m^2}\int
_0^t dt_1 \int _0^t dt_2 \ \langle \mbox{\boldmath$p$}(t_1)\cdot
\mbox{\boldmath$p$}(t_2) \rangle
\nonumber \\
&=& \frac{1}{m^2} \int_0^t dt_1 \int _{0}^{t} dt_2 \ \langle p(t_1) p(t_2) \cos \theta
\rangle 
\end{eqnarray}
where $\theta$ is an angle between $\mbox{\boldmath$p$}(t_1)$ and
$\mbox{\boldmath$p$}(t_2)$. We recall that when the force is the gradient of
the potential, we have $D_{xx}\lll D_{yy}$ for $p\gg p_0$ implying
that the correlation of the angle vanishes much more rapidly than
the correlation of the magnitude of the momentum. We can, therefore,
perform the averaging in (\ref{eq: 5.1}) by first integrating over the 
correlation function of the angular variable, with the momentum held
fixed, and then finally performing the averaging over fluctuations of the 
momentum. 

In the two-dimensional case the probability density $P(\theta,t)$ of $\theta $ satisfies the
diffusion equation on a circle with the initial condition
$P(\theta,0)=\delta(\theta)$. The solution is Gaussian:
\begin{eqnarray}
\label{eq: 5.2}
P(\theta,t)&=&\frac{1}{2\sqrt{\pi
\mathcal{D}t}} \nonumber \\ 
&\times& {\rm exp}\left(
-\frac{\theta ^2}{4\mathcal{D}t}\right),
\end{eqnarray}
where $\mathcal{D}=D_1p_0/p^3(t_1)$. Using this probability
density we calculate the expectation value
\begin{equation}
\label{eq: 5.3}
\langle \cos \theta \rangle = {\rm
exp}\left(-\mathcal{D}|t_2-t_1|\right).
\end{equation}
We thus obtain from equations~(\ref{eq: 5.1}) and (\ref{eq: 5.3})
\begin{eqnarray}
\label{eq: 5.4}
\langle |\mbox{\boldmath$x$}(t)|^2\rangle &\sim& \frac{1}{m^2} \int _0^t
{\rm d}t_1
\int _0^t {\rm d}t_2 \int _0^{\infty} {\rm d}p \ \rho(p,t_1)p^3(t_1) \nonumber \\
&\times& {\rm exp}\left( -\mathcal{D}|t_2-t_1| \right).
\end{eqnarray}
Introducing a new variable $T=t_1-t_2$ we have
\begin{eqnarray}
\label{eq: 5.5}
\langle |\mbox{\boldmath$x$}(t)|^2\rangle  &\sim &   \frac{2}{m^2}\int _0^t
{\rm d}t_1 \int _0^t {\rm d}T\nonumber\\
&\times& \int _0^\infty {\rm d}p \ \rho(p,t_1)p^3(t_1) {\rm exp} \left( -\frac{D_1p_0}{p^3(t_1)}T \right) \nonumber \\
&=&\frac{2}{m^2D_1p_0}\int _0^t {\rm d}t_1 \int _0^\infty {\rm d}p \
\rho(p,t_1) p^6(t_1) \nonumber \\
&\times&\left[1-{\rm exp} \left(
-\frac{D_1p_0}{p^3(t_1)}t\right)\right]\ .
\end{eqnarray}
When the forcing is strong we have $D_1p_0t\gg p^3$ for
$t\gg \tau$, and therefore
\begin{eqnarray}
\label{eq: 5.6}
\langle |\mbox{\boldmath$x$}(t)|^2\rangle &=& \frac{2}{m^2D_1p_0}\int _0^t
{\rm d}t_1 \int _0^\infty {\rm d}p \ \rho(p,t_1) p^6(t_1)
 \nonumber\\
&=& \frac{2}{m^2D_1p_0}\int _0^t {\rm d}t_1 \langle p^5(t_1)\rangle\ .
\end{eqnarray}
Using equation~(\ref{eq: 4.4}) we obtain
\begin{equation}
\label{eq: 5.7}
\langle |\mbox{\boldmath$x$}(t)|^2\rangle  =
\frac{4D_3p_0^2}{5D_1m^2\gamma^2}(5\gamma t + {\rm e}^{-5\gamma t}
- 1 )\ .
\end{equation}
\begin{figure}[t]
\centerline{\includegraphics[width=8.0cm]{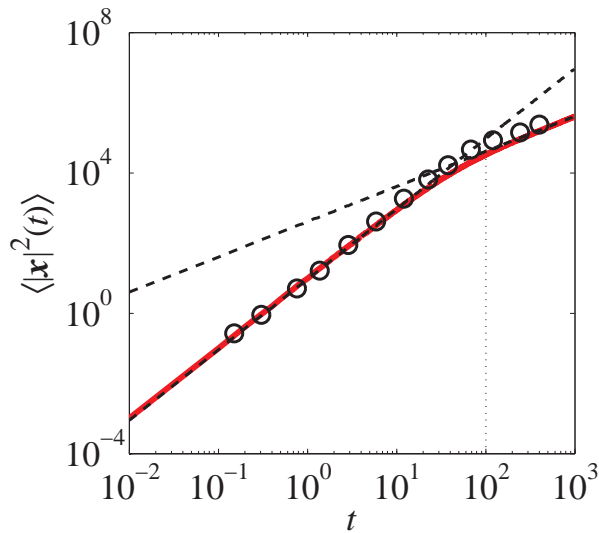}}
\caption{\label{fig: 4} Shows results for the
spatial diffusion in the two-dimensional potential force-field.
The results from the numerical simulation (circles) are compared with
equation~(\ref{eq: 5.7}) (solid line). Dashed lines show the slopes
$t^{2}$ and $t$ and dotted line indicates the time $\gamma^{-1}$. The parameters of the
simulation are the same as in figure~\ref{fig: 3}.}
\end{figure}

In the three-dimensional case the probability density of $\cos\theta$
can be found by considering the diffusion equation on a spherical surface,
with polar coordinates $(\theta,\phi)$, starting from the pole, $\theta=0$. 
The solution $P(\theta,\phi,t)$ of the diffusion equation may be expressed as a linear 
combination of spherical harmonics. Because the problem has a 
rotational symmetry, the solution is independent of the azimuthal 
angle $\phi$, and it may be written as:
\begin{equation}
\label{eq: 5.8}
P(\theta,\phi,t)=\sum _{l=0}^{\infty} A_l \ {\rm
exp}\left[-\frac{l(l+1)p_0D_1}{p^3}t\right]P_l(\cos\theta)
\end{equation}
where $P_l(z)$ is a Legendre polynomial of degree $l$. 
Using the orthogonality relations for Legendre polynomials,
 in view of the initial condition $\cos\theta=1$, 
we obtain  $A_l=(2l+1)/2$. Also, the quantity that we wish 
to average is itself a spherical harmonic: $\cos \theta=P_1(\cos\theta)$,
so that only the $l=1$ term in (\ref{eq: 5.8}) contributes to the correlation function. 
Hence we obtain
\begin{equation}
\label{eq: 5.9}
\langle \cos \theta (|t_2-t_1|) \rangle = {\rm
exp}\left(-3\mathcal{D}|t_2-t_1|\right),
\end{equation}
where $\mathcal{D}=p_0D_1/p^3$ is the same as in the
two-dimensional case. Using this angular correlation function we calculate 
\begin{eqnarray}
\label{eq: 5.10}
\langle |\mbox{\boldmath$x$}(t)|^2\rangle &\sim& \frac{1}{m^2} \int _0^t {\rm d}t_1
\int _0^t {\rm d}t_2 \int _0^\infty {\rm d}p \
\rho(p,t_1)p^4(t_1)
\nonumber \\ 
&\times&{\rm exp}\left(
-3\mathcal{D}|t_2-t_1| \right).
\end{eqnarray}
The evaluation of the integral using the probability density
(\ref{eq: 4.6}) yields
\begin{equation}
\langle |\mbox{\boldmath$x$}(t)|^2\rangle =
\frac{2D_3p_0^2}{5D_1m^2\gamma^2}(5\gamma t + {\rm e}^{-5\gamma t}
- 1 ).
\end{equation}
We find that in two- and three-dimensional cases $\langle |\mbox{\boldmath$x$}(t)|^2\rangle \sim
t^2$ at short times, so that the particle diffuses ballistically. The results are
consistent with a short-time asymptotic behaviour of the undamped
particle obtained in \cite{Ros92} for $d>1$. The long-time
behaviour is naturally diffusive, $\langle |\mbox{\boldmath$x$}(t)|^2\rangle
\sim t$. In Fig.~\ref{fig: 4} we show the comparison of the
analytical and numerical results for $\langle
|\mbox{\boldmath$x$}(t)|^2\rangle$ for the case of motion in the
two-dimensional potential force field, illustrating the short-time
ballistic diffusion.

\section{Summary}
\label{sec: 6}

We have investigated generalizations of two classical
models for diffusion of a particle accelerated by random forces.
We discussed a generalization of the classical
Ornstein-Uhlenbeck process where the force depends
on the position of the particle as well as time. We also
modified the Chandrasekhar-Rosenbluth model by considering motion
due to a short-range interaction potential. Although both models
are described by different microscopic equations of motion,
surprisingly, they have the same scaling of the diffusion
coefficients, leading to the same short-time asymptotic dynamics.

We solved the Fokker-Planck equation for the generalised
Ornstein-Uhlenbeck process exactly in two and three dimensions,
building upon our earlier analysis of the one-dimensional
case in \cite{Arv+05,Bez+06}. We have shown that this dynamics is characterised by
anomalous diffusion of the momentum, with the variance which
scales as $\langle p^2\rangle \sim t^{2/5}$. At long time, the
distribution of the momentum has been found to be non-Maxwellian.
The second moment of the displacement grows ballistically at short times, that
is $\langle x^2\rangle \sim t^2$, in accord with a surmise made 
by Rosenbluth for a closely related model \cite{Ros92}, and at 
long time a simple diffusive behaviour of
the displacement is recovered.

\end{document}